\documentclass[aps,preprint,preprintnumbers,amsmath,amssymb]{revtex4}
\usepackage{amssymb}
\usepackage{graphicx}
\usepackage{dcolumn}
\usepackage{bm}
\usepackage{color}

\makeatletter

\newcommand{\Rmnum}[1]{\expandafter\@slowromancap\romannumeral #1@}

\begin{document}

\title{Heavy Fermion Quantum Criticality and Destruction of the Kondo Effect in a Nickel Oxypnictide}

\author{Yongkang Luo$^{1,2}$, Leonid Pourovskii$^{3,4}$, S. E. Rowley$^{2}$, Yuke Li$^{5,1}$, Chunmu Feng$^{1}$, Antoine Georges$^{3}$, Jianhui Dai$^{5,1}$, Guanghan Cao$^{1}$, Zhu'an Xu$^{1}$\footnote[1]{Electronic address: zhuan@zju.edu.cn}, Qimiao Si$^{6}$, and N. P. Ong$^{2}$}

\address{$^1$Department of Physics and State Key Laboratory of Silicon Materials, Zhejiang University, Hangzhou 310027, P. R. China,}
\address{$^2$Department of Physics, Princeton University, Princeton, New Jersey 08544, U.S.A,}
\address{$^3$Centre de Physique Th\'{e}orique, \'{E}cole Polytechnique, CNRS, 91128 Palaiseau Cedex, France,}
\address{$^4$Swedish e-science Research Centre (SeRC), Department of Physics, Chemistry and Biology (IFM), Link\"oping University, Link\"oping, Sweden}
\address{$^5$Department of Physics, Hangzhou Normal University, Hangzhou 310036, P. R. China, }
\address{$^6$Department of Physics and Astronomy, Rice University, Houston, TX 77005, USA}

\date{\today}

\maketitle

\textbf{A quantum critical point arises at a continuous transformation between distinct phases of matter at zero temperature. Studies in antiferromagnetic heavy fermion materials have revealed that quantum criticality has several classes, with an unconventional type that involves a critical destruction of the Kondo entanglement\cite{Coleman-QC,Si-QCP}. In order to understand such varieties, it is important to extend the materials basis beyond the usual setting of intermetallic compounds. Here we show that a nickel oxypnictide, CeNiAsO, displays a heavy-fermion antiferromagnetic quantum critical point as a function of either pressure or P/As substitution. At the quantum critical point, non-Fermi liquid behavior appears, which is accompanied by a divergent effective carrier mass. Across the quantum critical point, the low-temperature Hall coefficient undergoes a rapid sign change, suggesting a sudden jump of the Fermi surface and a destruction of the Kondo effect \cite{Paschen-QCPHall,Custers-Ce3Pd20Si6}. Our results imply that the enormous materials basis for the oxypnictides, which has been so crucial to the search for high temperature superconductivity, will also play a vital role in the effort to establish the universality classes of quantum criticality in strongly correlated electron systems.}

The iron-based pnictides have not only revived the study of high temperature superconductivity \cite{Hosono-LaOF,WangNL-CeOF}, but also provided
a new material class to investigating quantum criticality\cite{Kasahara-Ba122_P} and heavy-fermion behavior. A canonical system for the latter is the 1111-type rare-earth iron pnictide CeFeAsO\cite{Geibel-CeP,XuZA-CeFeAsPO,Ishida-CeRuFePO}. The heavy-fermion behavior crucially depends on the delicate interplay between the $3d$- and $4f$-electrons\cite{DaiZhuSi}; indeed, the presence of the $3d$-electron antiferromagnetic (AFM) order in the FeAs layer complicates the study of the $4f$ electrons. We therefore turn to CeNiAsO, the Ni counterpart of the parent iron pnictides, in which the $3d$ electrons are shown to be magnetically disordered based on the measurements of entropy and magnetic susceptibility\cite{XuZA-CeNiAsO}.

The compound CeNiAsO is homology to the well-studied CeFeAsO\cite{WangNL-CeOF} of the ZrCuSiAs crystalline structure. The absence of AFM order on the Ni sublattice was confirmed both theoretically and experimentally\cite{XuG-LaOMAs,Ronning-Ni2X2}. There exists a substantial hybridization between the Ce-$4f$ and Ni-$3d$ electrons, which results in an enhanced Sommerfeld coefficient $\gamma_0$$\sim$ 203 mJ/mol$\cdot$K$^{2}$ (as estimated from the paramagnetic state) and a relatively high Kondo scale $T_K$$\sim$15 K\cite{XuZA-CeNiAsO}. These properties make CeNiAsO an ideal candidate to reveal the heavy-fermion properties of the $4f$-electrons.

Figure 1 shows the temperature dependence of resistivity $\rho(T)$ at various pressures in CeNiAsO. The resistivity data at ambient pressure contain two prominent features: a hump around 120 K and a sharp decrease below 10 K. Thermopower and specific heat measurements\cite{XuZA-CeNiAsO} suggested that the hump around 120 K is due to the crystalline electric field (CEF) effect, while the sharp decrease below 10 K is caused by the reduction of the spin flip scattering after the formation of the long range Ce$^{3+}$ AFM ordering. Further analysis indicated two separate AFM transitions, with the transition temperatures $T_{N1}$=9.3 K and $T_{N2}$=7.3 K, respectively (see also Figure 1c). Under pressure, both AFM transitions are suppressed (inset of Figure 1a). To demonstrate this evolution more clearly, we plot $d\rho/dT$ vs. $T$ curves for several representative pressures in Figure 1c. At first both $T_{N1}$ and $T_{N2}$ are reduced under pressure. When pressure increases to above 4 kbar, the peak corresponding to $T_{N1}$ in $d\rho/dT$ was hardly seen, while $T_{N2}$ is continuously suppressed at the pressure $p$$\sim$6.7 kbar [see Figure 1d and Figure S1 in Supplementary Information (SI)]. A nearly-linear $\rho(T)$ dependence was observed at this pressure, manifesting non-Fermi-liquid (NFL) behavior (see Figure 1d and Figure 2d). At even higher pressures, the resistivity exhibits typical Kondo lattice behavior (Figure 1b), with $\rho(T)$ decreasing slowly at high temperature and dropping much faster below $T_{0}$, the onset temperature of Kondo coherence as illustrated in the inset of Figure 1b. Pressure increases the hybridization matrix element $V_{fc}$, leading to an enhancement of $T_0$ (see SI). A Fermi liquid (FL)-like $\rho$ vs. $T$ dependence is observed well below $T_{0}$ (see Figure 1e-f). No superconductivity can be identified for all measured pressures and temperatures down to 0.37 K. The quantum critical point (QCP) is defined as the pressure where $T_{N2}$ extrapolates to zero from below. The pressure dependence of $T_{N1}$, $T_{N2}$, and $T_{FL}$ are shown in Figure 2a, which leads to $p_c$$\approx$6.5 kbar. Also shown in Figure 2a is $T_0$ which increases with pressure.

We can now analyze the evolution of the electrical resistivity across the QCP. We fit the low-temperature resistivity
$\rho(T)$ to the formula $\rho$=$\rho_0$+$\Delta\rho$=$\rho_0$+$A T^n$, where $\rho_0$ is the residual resistivity.
A local fitting on the ($p$,$T$) phase diagram leads to $n(p,T)$=$d(\ln\Delta\rho)/d(\ln T)$ which is shown
in Figure 2a. This allows us to define $T_{FL}$, below which FL behavior is present with $n$$\approx$2
(taken as 1.8$\leq$$n$$\leq$2.2). The approach to QCP is also manifested in the pressure dependence
of $\rho_0$, which is peaked at 5.6 kbar, close to $p_c$ (see Figure 2b). We note that $T_{N1}$ is suppressed  already at 4 kbar, and this might broaden the peak in $\rho_0$ vs. pressure and shift the peak position to somewhat below the critical pressure $p_c$. Finally, our analysis also leads to the $A$ coefficient according to the $\Delta\rho$=$A T^2$ fit in the $T$$\rightarrow$0 limit. As presented in Figure 2c, this $A$ coefficient shows a pronounced peak at $p_c$. Quite strikingly, its value near the QCP is enhanced by more
than three orders of magnitude compared to that measured far away from the QCP. We also plot $A T_0^2$ vs.
$(p-p_c)$ in the inset of Figure 2c. The scaling form of $A T_0^2$$\propto$$1/(p-p_c)^{\alpha}$ with $\alpha$=0.8
implies that the $A$ coefficient increases much faster than $T_0$ approaching $p_c$; this is to be expected,
given that $T_0$ stays nonzero across $p_c$ even though it shows a significant $p$-dependence at higher
pressures. The $A$ coefficient is related to the effective carrier mass\cite{Kadowaki-Woods} through
$A$$\sim$$(m^*/m_0)^2$. The divergence of $A$, therefore, signifies a divergence of the effective carrier mass.
The latter indicates a localized-delocalized transition of the 4$f$-electrons\cite{Si-Nature}.

To obtain further information about the quantum phase transition, we also synthesized and investigated phosphorus doped CeNiAsO. Since the ionic radius of P$^{3-}$ is less than As$^{3-}$, this isovalent substitution serves as a positive chemical pressure\cite{WangC-LaP}. The sample quality of CeNiAs$_{1-x}$P$_x$O was checked by powder x-ray diffraction (XRD) (see Figure S2 in SI). The lattice parameters, both $a$ and $c$, decrease monotonically with P doping, confirming the application of such chemical pressure. In general, this effect of the chemical pressure is consistent with that of the hydrostatic pressure (see Figure S3 in SI). The critical doping concentration is $x_c$$\approx$0.4. Unlike applying pressure to the pure CeNiAsO, the P-for-As doping also allows the measurement of the specific heat, which is displayed in Figure 3. In the $x$=0 sample, the specific heat reveals two phase transitions below 10 K confirming the two AFM transitions determined by resistivity measurements\cite{XuZA-CeNiAsO}. For $x$=0.2, however, only one peak can be identified, and the peak position is much lower than that of $x$=0; again, this  is consistent with the resistivity measurement (see Figure S3c in SI). At the critical point $x$=0.4, no trace of phase transition can be found down to 0.5 K; moreover, the electronic contribution $\gamma (T)$$\equiv$$C_{el}/T$ derived by subtracting a $\beta T^2$ term of phonon contribution increases logarithmically below 10 K as shown in the inset of Figure 3. This divergent $\gamma(T)$ provides further evidence for a divergent quasiparticle mass at the QCP. A tiny saturation trend of $C_{el}/T$ can be seen below 3 K, and the Sommerfeld coefficient $\gamma_0$ is estimated to be above 700 mJ/(mol$\cdot$ K$^2$); this likely signifies that the critical concentration is very close to, but slightly away, from $x$=0.4. Far away from the critical point, both CeNiAs$_{0.4}$P$_{0.6}$O and CeNiPO behave like a Fermi liquid with moderately enhanced effective mass, while for $x$=0.45, the broad peak centered at 100 K$^2$ in the $C_{el}/T$ vs. $T^2$ plot of Figure 3 ({\it i.e.}, $T$$\simeq$10 K) is a clear signature of the Kondo effect. The anomalies around 35 K$^2$
in the $x$=0.5, 0.6 and 1.0 samples are attributed to tiny amounts of the magnetic impurities Ce$_2$O$_3$ \cite{Huntelaar-Ce2O3}, since the peak position does not change with P doping; such an impurity phase is confirmed by the XRD patterns (see those $*$ marks in Figure S2). We plot $C/T$ at 2 K in Figure S3e
as a function of P doping $x$. The peak in $C_{2K}/T$ is consistent with that in the $x$-dependence of the $A^{0.5}$ coefficient obtained
from resistivity measurement, which is also shown in Figure S3e. We conclude that the chemical pressure on CeNiAsO also results in a heavy fermion QCP.

We have also measured the magnetic susceptibility of CeNiAs$_{1-x}$P$_x$O, which is displayed in Figure 4. Since the 3d electrons of the Ni ions do not order magnetically in the nickel based pnictides\cite{Ronning-Ni2X2}, the observed magnetic response should be dominated by trivalent Ce ions. Several properties are observed. First, the peak associated with the AFM transition shifts to lower temperature with increasing $x$, and disappears at around $x$=0.4 (inset of Figure 4a). Second, $\chi(T)$, in the paramagnetic state for $x$$<$0.4, can be fit to the Curie-Weiss form, $\chi(T)$=$\chi_0$$+$$C/(T$$-$$\theta_W)$, where $\chi_0$ is a temperature-independent constant and $\theta_W$ is the Weiss temperature. The derived effective moment does not show much variation with $x$ for $x$$<$0.4, which is consistent with the Kondo temperature scale $T_0$ being essentially $x$-independent in this $x$ range. It is about 2.24 $\mu_B$ ($\mu_B$ is Bohr magneton), which is close to but slightly smaller than that of the free Ce$^{3+}$ ion, 2.54 $\mu_{B}$; the deviation is ascribed to CEF effect, as is the case in CeNiAsO\cite{XuZA-CeNiAsO}. This provides further evidence that the Ce-$4f$ electrons in CeNiAs$_{1-x}$P$_x$O appear as local moments, which in turn  form a magnetically-ordered ground state, at $x$$<$0.4. Third, when $x$$\geq$0.4, $\chi(T)$ becomes more and more temperature independent which means a reduction of the local magnetic moment, and this is consistent with a delocalization of the Ce-$4f$ electrons due to the enhanced Kondo coupling. The evolution of the Ce-$4f$ magnetic moment is also supported by isothermal magnetization measurement, as shown in Figure S4 in SI. Forth, a coherent Kondo screening is present over a broad temperature range, and it decreases the magnitude of susceptibility. For example, in CeNiPO (see Figure 4b), $\chi(T)$ follows the Curie-Weiss law for $T$$>$300 K and $T$$<$100 K, but violates it at the intermediate temperatures. In this broad temperature range 100 K$<$$T$$<$300 K, $\chi(T)$ decreases upon cooling down, and the effective moment is reduced from $\mu_{eff}^h$=1.30 $\mu_B$ to $\mu_{eff}^l$=0.21 $\mu_B$. This observation is consistent with the resistivity measurements; a characteristic temperature $T_0$ can be defined as shown in inset of Figure 4b. For CeNiPO, we find $T_0$=200 K, which is comparable with the value (207 K) extracted as twice the temperature at which the electronic entropy reaches 0.4$R\ln2$\cite{Si-QCP}. The divergent increase in $\chi(T)$ at low temperatures originates from the magnetic impurities discussed earlier. We plot $T_0$ on the phase diagram in Figure 2a and Figure S3e for hydrostatic pressure and chemical pressure, respectively.

One striking feature here is the substantial variation of $T_0$ as a function of both physical pressure and chemical pressure. We have studied this feature through $ab$-initio calculations using local density approximations in combination with dynamical mean field theory (LDA+DMFT) \cite{kotliar2006,aichhorn2009,Wien2K} on CeNiAsO and CeNiPO at $T$=290 K and 12 K (more details in SI). A Kondo peak formed by renormalized Ce-$4f$ states is observed in CeNiPO in the vicinity of the Fermi level $E_F$ already at $T$=290 K, and is sharpened at lower temperatures. By contrast, it is absent in CeNiAsO even at $T$=12 K (Figure S5). This difference in the high temperature behavior between CeNiAsO and CeNiPO is due to a stronger $3d$-$4f$ hybridization and $4f$ level position being closer to $E_F$ in the latter. The calculated Kondo scales\cite{Gunn1983} $T_K$=15(527) K for CeNiAsO (CeNiPO), respectively, are in good agreement with the experimental measurement, and is compatible with the rapidly increasing $T_0$ with pressure.

Two types of QCPs have been advanced for AFM heavy fermion metals. One type corresponds to a $T$=0 SDW transition, with the quantum criticality described in terms of fluctuations of the SDW order parameter\cite{Hertz-QCP}. Another type invokes a continuous destruction of the Kondo effect at the AFM transition,
and is accompanied by a sudden jump of the Fermi-surface across the QCP\cite{Si-Nature,Coleman-FLheavy}. Experimentally, YbRh$_2$Si$_2$ has been evidenced as displaying the Kondo-destruction QCP\cite{Paschen-QCPHall,Gegenwart-YbRh2Si22,Friedemann10}.

To study the evolution of the Fermi surface, we have measured the Hall effect at 0.38 K under various pressures; the data are shown in Figure 2e. Through a very rapid change across $p_c$, $R_H$ goes from being negative at $p$$<$$p_c$ to being positive at $p$$>$$p_c$, suggesting that a drastic change in FS takes place as the QCP crossed. While future experiments are needed to ascertain the temperature dependence of the crossover width especially
at even  lower temperatures\cite{Paschen-QCPHall}, we stress that the sign change occurs over a very narrow
pressure that is almost the resolution limit of our pressure cell equipment.
Moreover, the quantum phase transition is evidenced to be continuous which implies the existence of QCP (see SI).
In the chemical pressure case, the $R_H$ jump near the QCP is even more abrupt (see inset of Figure S4b).
The consistency between the $R_H$ jump observed as a function of both pressure and chemical substitution
is to be contrasted with the case of V-doped Cr, where the absence of divergent $A$ or $\gamma$ coefficients
makes it necessary for very fine steps of tuning by pressure to reveal the critical behavior of the Hall coefficient\cite{Lee-CrHall}. At the same time, our results reveal in CeNiAsO a Hall-coefficient jump across a QCP
in the absence of any magnetic field. When combined with the field-induced cases such as realized in YbRh$_2$Si$_2$
\cite{Paschen-QCPHall,Gegenwart-YbRh2Si22,Friedemann10}, our result provides evidence for the robustness
of the effect. It also rules out the mechanism of Zeeman-driven Lifshitz transition \cite{Hackl-Lifshitz}. Because CeNiAsO and CeNiPO contain multiple bands,
it is instructive to gain a microscopic interpretation of the sign change in $R_H$ through electronic-structure calculations. We have therefore used the DMFT technique to calculate the FS topology as depicted in Figure S6 in SI. Indeed, the calculated FS of CeNiAsO is very similar to that of LaNiAsO\cite{XuG-LaOMAs}, which contains no $4f$-electrons. In contrast, the FS of CeNiPO is drastically modified by temperature due to a gradual formation of the heavy-electron Ce-$4f$ band in the vicinity of Fermi level. One should also notice that the total number of electrons enclosed by FS has been significantly enlarged from CeNiAsO to CeNiPO, which is a consequence of participation into the $DOS$ of Ce-$4f$ electrons. In addition, a three-dimensional hole pocket develops centering at the $Z$ point (Figure S6f), which makes a positive contribution to the Hall coefficient. The sign change we have observed in the Hall coefficient,
combined with the divergent quasi-particle effective mass, is therefore consistent with a sudden change of Fermi surface from small to large as pressure is increased across $p_c$.

To summarize, the absence of Ni-$3d$ correlated magnetism has allowed us to unambiguously identify a magnetic quantum critical point in a nickel oxypnictide under both hydrostatic and chemical pressure. Near the quantum critical point, the system displays non-Fermi-liquid behavior, a divergent effective carrier mass and a sudden sign change of the Hall coefficient. These results provide the first clear evidence for Kondo destruction in an oxypnictide, thereby extending the unconventional local heavy-fermion quantum criticality to a new category of materials. More generally, our study points to the prospect that the oxypnictides will provide a large materials basis to understand the universality classes of quantum criticality.\\

\textbf{METHODS}

--------------------------------------------------------

\textbf{\Rmnum{1}. Sample synthesis}

Poly-crystalline CeNiAs$_{1-x}$P$_{x}$O ($x$=0, 0.1, 0.2, 0.3, 0.35, 0.4, 0.45, 0.5, 0.6, 0.8, and 1.0) samples were synthesized by solid state reaction. Ce, Ni, As, and CeO$_{2}$ of high purity ($\geq$99.95\%, Alfa Aesar) were used as starting materials. Firstly, CeAs (CeP) was pre-synthesized by reacting Ce discs and As (P) powders at 1320 K for 72 h. NiAs (NiP) was pre-synthesized by reacting Ni and As (P) powders at 970 K for 20 h. Secondly, powders of CeAs, CeP, CeO$_{2}$, Ni, NiAs and NiP were weighted in stoichiometric ratio, thoroughly ground, and pressed into a pellet under a pressure of 600 MPa in an Argon filled glove box. The pellet was packed in an alumina crucible and sealed into an evacuated quartz tube, which was then slowly heated to 1450 K and kept at that temperature for 40 h. All the samples were checked by powder X-ray diffraction (XRD) which was performed at room temperature using a D/Max-rA diffractometer with Cu-$K_{\alpha}$ radiation and a graphite monochromator.

\textbf{\Rmnum{2}. Measurement}

Piston-cylinder pressure cell was used in the high pressure experiment, during which highly pure Pb was used as the manometer. The pressure was determined by the pressure dependent $T_c(p)$ of Pb\cite{lead-Tc}, and was double checked by resistance ratio $R(p)/R(0)$ at room temperature\cite{R(p)/R(0)}. Daphne 7373 oil was used as the pressure fluid, and hydrostatic pressure up to 26 kbar was applied in the experiment. He-3 refrigerater was used to get the lowest temperature 0.37 K. Hall coefficient was derived by sweeping magnetic field from -5 T to 5 T. The dc magnetization measurement for CeNiAs$_{1-x}$P$_{x}$O was carried out in a Quantum Design magnetic property measurement system (MPMS-5) in zero-field-cooling (ZFC) and field-cooling (FC) protocals under a magnetic field $H$ of 1000 Oe, while specific heat was measured by heat pulse relaxation method in Quantum Design physical property measurement system (PPMS-9).

\textbf{\Rmnum{3}.Theoretical calculations}

The details of theoretical calculations are displayed in Supplementary Information.

\begin{acknowledgments}

This work was supported by the National Basic Research Program of China (Grant Nos.  2011CBA00103 and 2012CB927404),  the National Science Foundation of China
(Grant Nos. 11190023, 11174247, 10934005 and 11274084), the NSF of Zhejiang Province
(No. Z6110033), the Fundamental Research Funds for the Central Universities of China,
the National Science Foundation under grant Nos DMR 0819860 and DMR-1309531,
the Nano Electronics Research Corporation (Award 2010- NE-2010G),
and the Robert A.\ Welch Foundation Grant No.C-1411. Y. Luo would like to acknowledge
a scholarship granted by China Scholarship Council (CSC-2010632081).
The DMFT calculations have been performed using computational facilities
of the Swedish National Infrastructure for Computing (SNIC) under projects 003-11-1
and 001-11-125.

\end{acknowledgments}

\textbf{Author contributions:}
Y. Luo, N. P. Ong, Q. Si and Z. A. Xu designed the research. Y. Luo synthesized the samples, and performed most of the measurements. L. Pourovskii and A. Georges made the first-principles calculations. N. P. Ong and S. E. Rowley provided important equipments and helpful discussions.  S. E. Rowley, C. Feng and Y. Li did some of the measurements. G. Cao, J. Dai, Y. Luo, L. Pourovskii, Q. Si, and Z. A. Xu discussed the data, interpreted the results, and wrote the paper.

\textbf{Author information:}
The authors declare no competing financial interests. Correspondence and requests for materials should be addressed to Zhu'an Xu (zhuan@zju.edu.cn).


\newpage
\textbf{Figure legend:}

\begin{figure}[!h]
\includegraphics[width=16cm]{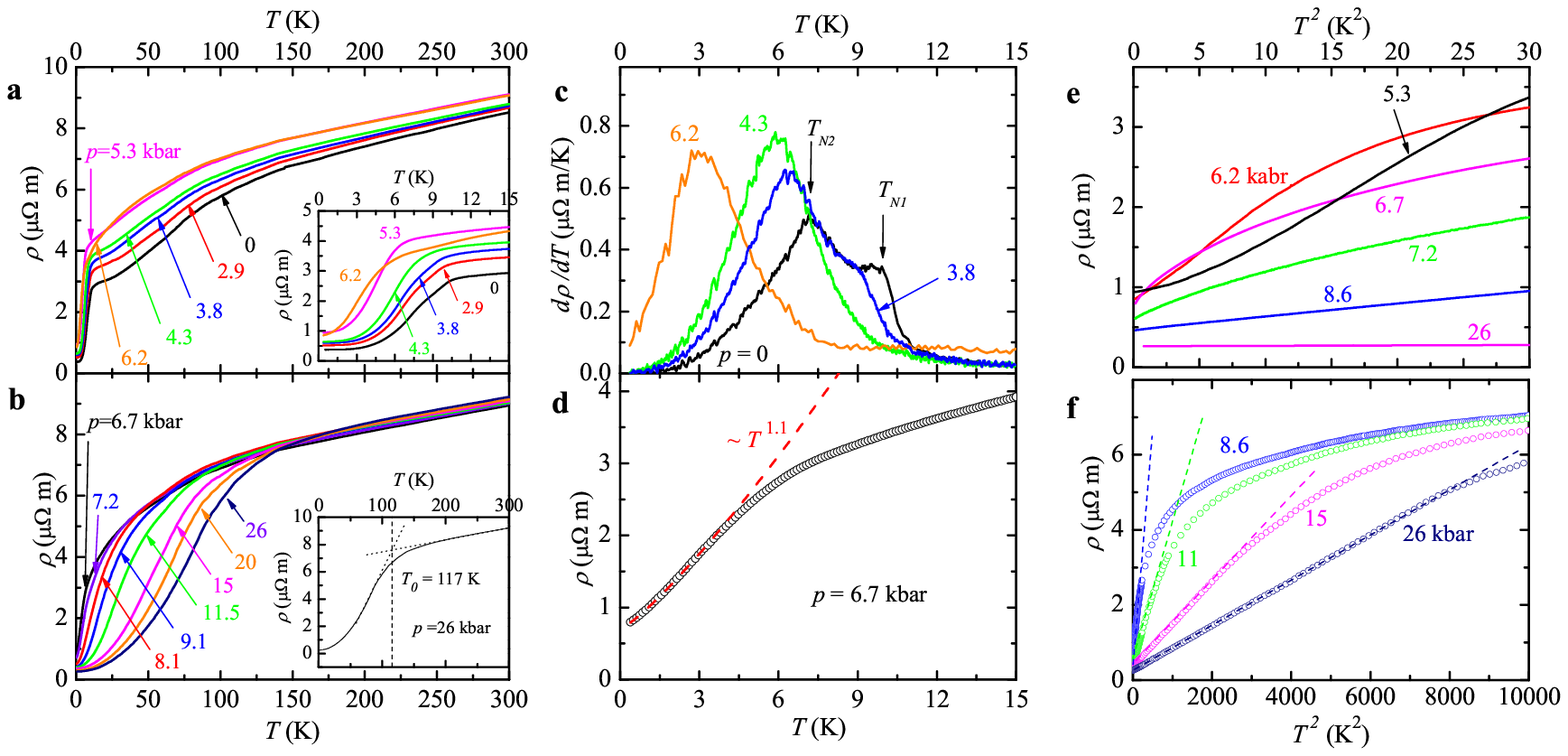}
\label{Fig1}
\end{figure}
\textbf{Figure 1 $|$ Resistivity vs. temperature in CeNiAsO under different hydrostatic pressures.}
\textbf{a-b}, profiles of $\rho(T)$ for $p$$\leq$6.2 kbar and $p$$\geq$6.7 kbar, respectively. The inset of \textbf{a} is the enlarged plot of the $\rho(T)$ curves at low temperatures $T$$\leq$15K, while the inset of \textbf{b} illustrates the definition of $T_0$ which characterizes the strength of Kondo interaction. \textbf{c}, $d\rho/dT$ curves for $p$=0,3.8,4.3,6.2 kbars, which show the suppression of the two AFM transitions under pressure. \textbf{d}, non-FL behavior observed at $p$=6.7 kbar with $\Delta\rho$$\propto$$T^{1.1}$. \textbf{e-f}, resistivity plotted vs. $T^2$, demonstrating the FL behavior at low temperatures for $p$$>$$p_c$, as well as the evolution of the $T^2$ slope ($A$) with pressure.\\

\begin{figure}[!h]
\includegraphics[width=15cm]{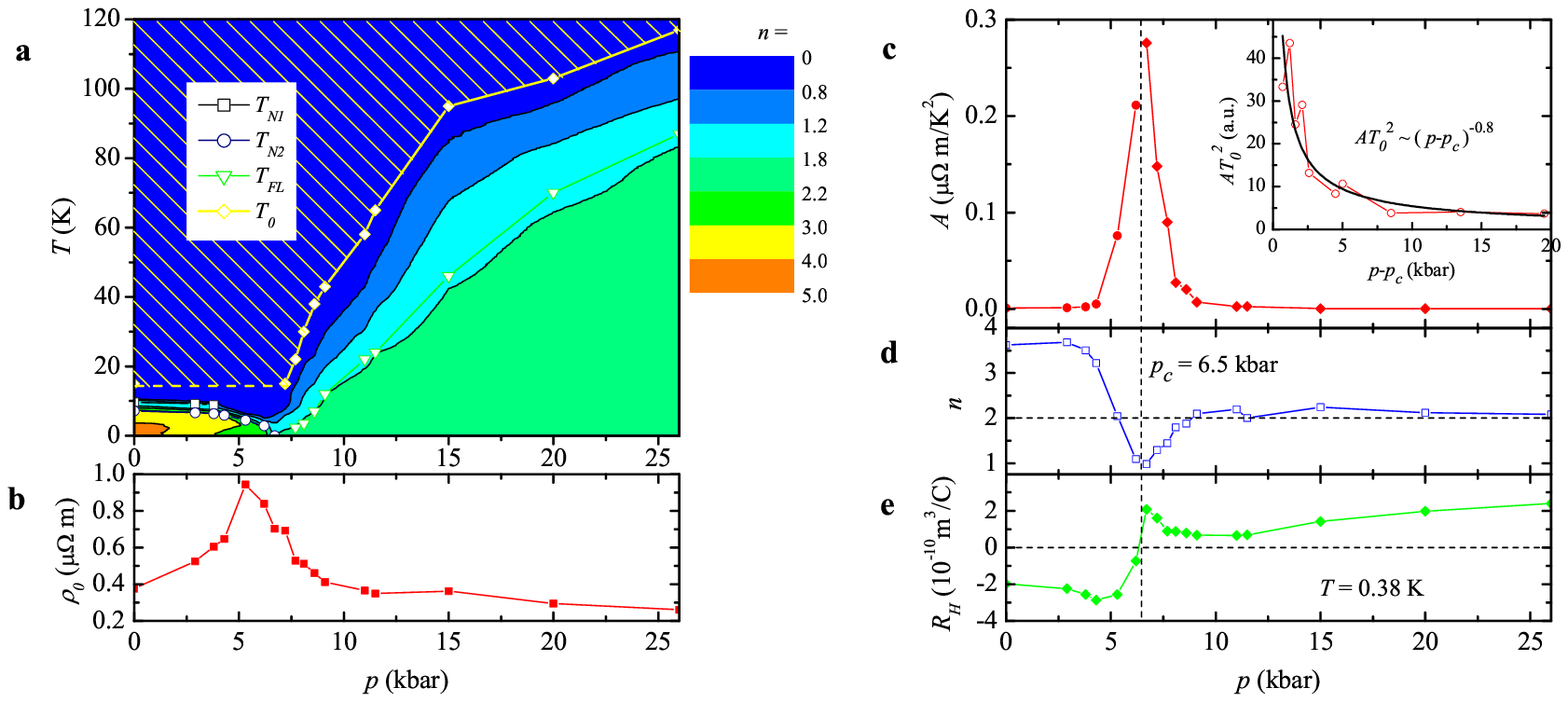}
\label{Fig2}
\end{figure}
\textbf{Figure 2 $|$ Phase diagram of CeNiAsO under pressure.}
\textbf{a}, contour plot of $n(p,T)$, where $n$ is the power of $\Delta \rho$=$A T^n$. The green area at the low right corner displays FL behavior which is defined by 1.8$\leq$$n$$\leq$2.2. Non-FL region, colored with light blue, is clearly seen around the QCP $p_c$=6.5 kbar. The green triangles, labeled by $T_{FL}$, was obtained from the starting point above which $\rho(T)$ deviates from the $T^2$ law as temperature is raised. The yellow diamonds signify the  onset of Kondo coherence estimated from resistivity (see inset of Figure 1b). For $p$$<$$p_c$, the dashed yellow line schematically shows a finite $T_0$ (which is equal to 15 K at ambient pressure \cite{XuZA-CeNiAsO}, but cannot be determined directly at nonzero pressure up to $p_c$). The shaded region stands for the incoherent Kondo scattering regime. The quantum critical behavior occurs at temperatures below $T_0$, as is also seen by the temperature dependence of the resistivity at the critical pressure, shown in Figure 1d (and, in the case of P-doping, the temperature dependence of the Sommerfeld coefficient at the critical doping concentration, plotted in the inset to Figure 3).
\textbf{b}, the residual resistivity $\rho_0$ as a function of pressure. The peak in $\rho_0(p)$ centered at 5.6 kbar signifies strong magnetic fluctuation near QCP. \textbf{c}, the pressure dependence of the $A$ coefficient, which tends to diverge near $p_c$. The inset of \textbf{c} shows $A$ multiplied by $T_0^2$, which obeys a power law $A T_0^2$$\propto$$(p-p_c)^{-0.8}$. \textbf{d} shows $p$ dependent resistivity exponent $n$ from $\rho(T)$ at $T$$\leq$5 K. \textbf{e}, the pressure dependence of the Hall coefficient, measured at 0.38 K, showing a rapid sign change near $p_c$=6.5 kbar.\\

\begin{figure}[!h]
\includegraphics[width=10cm]{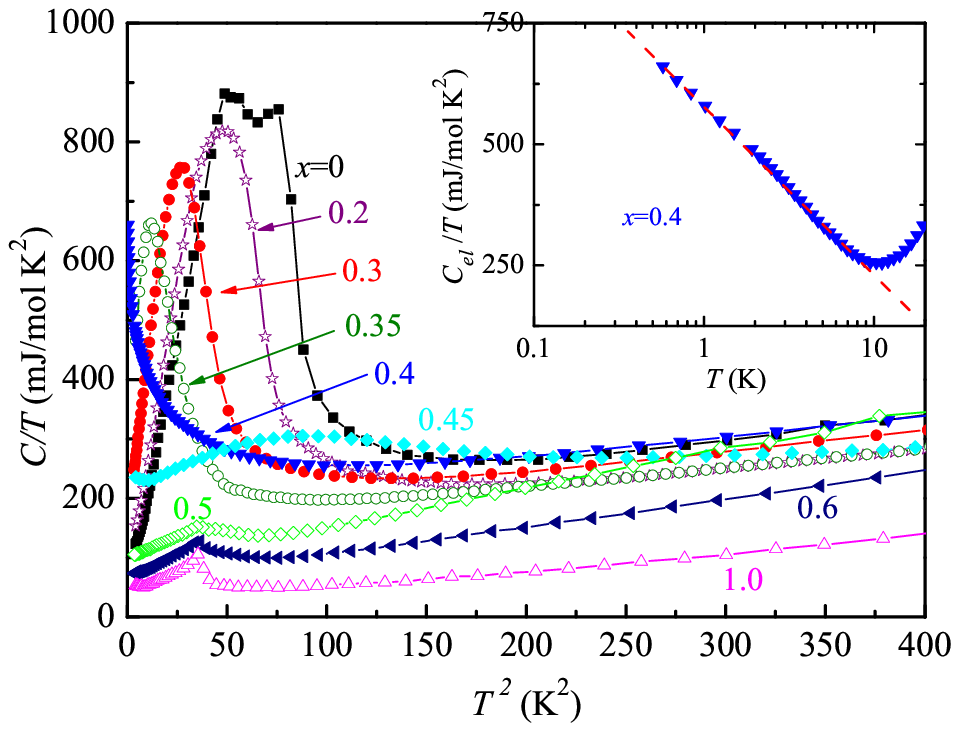}
\label{Fig3}
\end{figure}
\textbf{Figure 3 $|$ Specific heat of CeNiAs$_{1-x}$P$_x$O.}
Mainframe shows temperature dependent specific heat of CeNiAs$_{1-x}$P$_x$O. The $x$=0 sample exhibits two phase transitions below 10 K, while only one peak is observed in the $x$=0.2 sample. At the critical concentration, $x$=0.4, no trace of phase transition is observed down to 0.5 K. For $x$=0.6 and 1.0, it behaves like a FL metal. Kondo coherence can be identified in $x$=0.45. The small peaks around 6 K are
attributed to small quantities of the magnetic impurities Ce$_2$O$_3$\cite{Huntelaar-Ce2O3}. The inset shows the electronic contribution $C_{el}/T$ vs. $\log(T)$ for $x$=0.4.\\

\begin{figure}[!h]
\includegraphics[width=10cm]{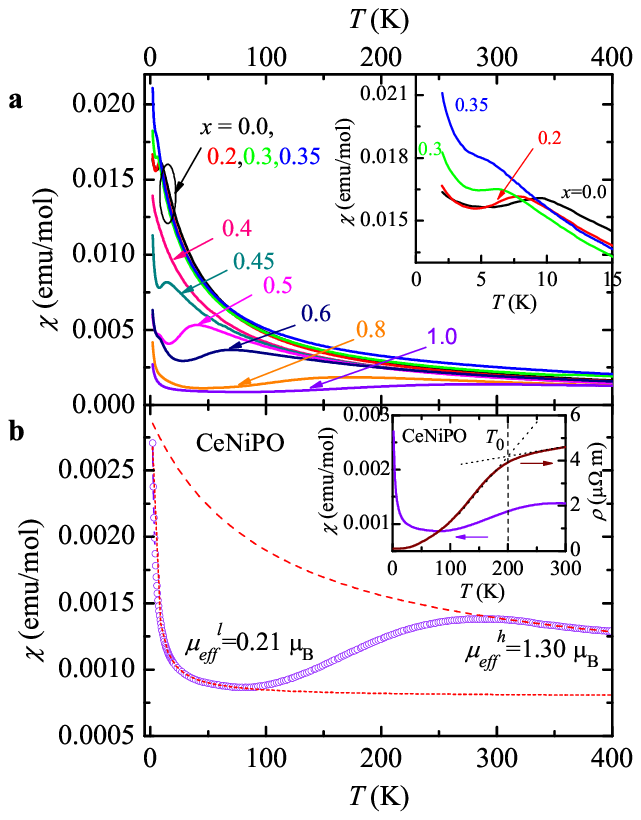}
\label{Fig4}
\end{figure}
\textbf{Figure 4 $|$ Magnetic property of P doped CeNiAsO.}
\textbf{a}, temperature dependent susceptibility of CeNiAs$_{1-x}$P$_x$O (0$\leq$$x$$\leq$1). The inset shows the evolution of the AFM transition with P doping. \textbf{b}, expanded view of $\chi(T)$ for CeNiPO, showing a drop in a broad temperature range centered around 200 K. Curie-Weiss fits of the data at high temperatures ($T$$>$300 K) and low temperatures ($T$$<$100 K) yield $\mu_{eff}^h$=1.30 $\mu_B$ and $\mu_{eff}^l$=0.21 $\mu_B$, respectively (see main text). Inset of \textbf{b} associates this drop with Kondo screening, while also specifying the value for $T_0$.\\


\begin{thebibliography}{00}

\bibitem{Coleman-QC}
Coleman, P. \& Schofield, A. J. Quantum criticality, \emph{Nature} \textbf{433}, 226-229 (2005).
\bibitem{Si-QCP}
Gegenwart, P., Si, Q. \& Steglich, F. Quantum criticality in heavy-fermion metals. \emph{Nature Physics} \textbf{4}, 186-197 (2008).
\bibitem{Custers-Ce3Pd20Si6}
Custers, J. \emph{et al.} Destruction of the Kondo effect in the cubic heavy-fermion compound Ce$_3$Pd$_{20}$Si$_6$, \emph{Nature Mater.} \textbf{11}, 189-194 (2012).
\bibitem{Paschen-QCPHall}
Paschen, S. \emph{et al.} Hall-effect evolution across a heavy-fermion quantum critical point. \emph{Nature} \textbf{432}, 881-885 (2004).
\bibitem{Hosono-LaOF}
Kamihara, Y., Watanabe, T., Hirano, M. \& Hosono, H. Iron-based layered superconductor La[O$_{1-x}$F$_x$]FeAs (x=0.05-0.12) with $T_c$=26 K. \emph{J. Am. Chem. Soc.} \textbf{130}, 3296-3297 (2008).
\bibitem{WangNL-CeOF}
Chen, G. F. \emph{et al.} Superconductivity at 41 K and Its Competition with Spin-Density-Wave Instability in Layered CeO$_{1-x}$F$_x$FeAs. \emph{Phys. Rev. Lett.} \textbf{100}, 247002 (2008).
\bibitem{Kasahara-Ba122_P}
Kasahara, S. \emph{et al.} Evolution from non-Fermi- to Fermi-liquid transport via isovalent doping in BaFe$_2$(As$_{1-x}$P$_x$)$_2$ superconductors. \emph{Phys. Rev. B} \textbf{81}, 184519 (2010).
\bibitem{Geibel-CeP}
Br\"{u}ning, E.M. \emph{et al.} CeFePO: A Heavy Fermion Metal with Ferromagnetic Correlations. \emph{Phys. Rev. Lett.} \textbf{101}, 117206 (2008).
\bibitem{XuZA-CeFeAsPO}
Luo, Y. \emph{et al.} Phase diagram of CeFeAs$_{1-x}$P$_x$O obtained from electrical resistivity, magnetization, and specific heat measurements. \emph{Phys. Rev. B} \textbf{81}, 134422 (2010).
\bibitem{Ishida-CeRuFePO}
Kitagawa, S., Ishida, K., Nakamura, T., Matoba, M. \& Kamihara, Y. Ferromagnetic Quantum Critical Point in Heavy-Fermion Iron Oxypnictide CeRu$_{1-x}$Fe$_x$PO, \emph{Phys. Rev. Lett.} \textbf{109}, 227004 (2012).
\bibitem{DaiZhuSi}
Dai, J., Zhu, J.-X., \& Si, Q. $f$-spin physics of rare-earth iron pnictides: Influence of $d$-electron antiferromagnetic order on the heavy fermion phase diagram. \emph{Phys. Rev. B} \textbf{80}, 020505(R) (2009).
\bibitem{XuZA-CeNiAsO}
Luo, Y. \emph{et al.} CeNiAsO: an antiferromagnetic dense Kondo lattice.  \emph{J. Phys.: Condens. Matter} \textbf{23}, 175701 (2011).
\bibitem{XuG-LaOMAs}
Xu, G. \emph{et al.} Doping-dependent Phase Diagram of LaO$M$As ($M$=V-Cu) and Electron-type Superconductivity near Ferromagnetic Instability. \emph{EPL} \textbf{82}, 67002 (2008).
\bibitem{Ronning-Ni2X2}
Ronning, F. \emph{et al.} Ni$_2$X$_2$ (X=pnictide, chalcogenide, or B) Based Superconductors. \emph{Physica C} \textbf{469}, 396-403 (2009).
\bibitem{Kadowaki-Woods}
Miyake, K., Matsuura, T. \& Varma, C. M. Relation between resistivity and effective mass in heavy-fermion and A15 compounds. \emph{Solid State Commun.} \textbf{71}, 1149-1153 (1989).
\bibitem{Si-Nature}
Si, Q., Rabello, S., Ingersent, K., \& Smith, L. Locally critical quantum phase transitions in strongly correlated metals. \emph{Nature} \textbf{413}, 804-808 (2001).
\bibitem{WangC-LaP}
Wang, C. \emph{et al.} Superconductivity in LaFeAs$_{1-x}$P$_{x}$O: Effect of chemical pressures and bond covalency. \emph{EPL} \textbf{86}, 47002 (2009).
\bibitem{Huntelaar-Ce2O3}
Huntelaar, M. E., Booij, A. S., Cordfunke, E. H. P., \& van der Laan R. R. The thermodynamic properties of Ce$_2$O$_3$(s) from $T\rightarrow $ 0 K to 1500 K. \emph{J. Chem. Thermodynamics} \textbf{32}, 465-482 (2000).
\bibitem{kotliar2006}
Kotliar, G. \emph{et al.} Electronic structure calculations with dynamical mean-field theory. \emph{Rev. Mod. Phys.} \textbf{78}, 865-951 (2006).
\bibitem{aichhorn2009}
Aichhorn, M. \emph{et al.} Dynamical mean-field theory within an augmented plane-wave framework: Assessing electronic correlations in the iron pnictide LaFeAsO. \emph{Phys. Rev. B} \textbf{80}, 085101 (2009).
\bibitem{Wien2K}
Blaha, P., Schwarz, K., Madsen, G., Kvasnicka, D. \& Luitz, J. \emph{WIEN2k, An augmented Plane Wave + Local Orbitals Program for Calculating Crystal Properties.} (Techn. Universitat Wien, Austria, 2001).
\bibitem{Gunn1983}
Gunnarsson, O. and Sch\"onhammer, K. Electron spectroscopies for Ce compounds in the impurity model. \emph{Phys. Rev. B} \textbf{28}, 4315-4341 (1983).
\bibitem{Hertz-QCP}
Hertz, J. A. Quantum critical phenomena. \emph{Phys. Rev. B} \textbf{14}, 1165-1184 (1976).
\bibitem{Coleman-FLheavy}
Coleman, P., P\'{e}pin, C., Si, Q. \& Ramazashvili, R. How do Fermi liquids get heavy and die? \emph{J. Phys.: Condens. Matter} \textbf{13}, 723-738 (2001).
\bibitem{Gegenwart-YbRh2Si22}
Gegenwart, P. \emph{et al.} Multiple energy scales at quantum critical point. \emph{Science} \textbf{315}, 969-971 (2007).
\bibitem{Friedemann10} Friedemann, S. \emph{et~al.} Fermi-surface collapse and dynamical scaling near a quantum critical point. \emph{Proc.\ Natl.\ Acad.\ Sci.\ USA} \textbf{107}, 14547-14551 (2010).
\bibitem{Lee-CrHall}
Lee, M., Husmann, A., Rosenbaum, T. F. \& Aeppli, G. High Resolution Study of Magnetic Ordering at Absolute Zero. \emph{Phys. Rev. Lett.} \textbf{92}, 187201 (2004).
\bibitem{Hackl-Lifshitz}
Hackl, A. \& Vojta, M. Zeeman-Driven Lifshitz Transition: A Model for the Experimentally Observed Fermi-Surface Reconstruction in YbRh$_2$Si$_2$. \emph{Phys. Rev. Lett.} \textbf{106}, 137002 (2011).
\bibitem{lead-Tc}
Clark, M. J. \& Smith, T. F. Pressure Dependence of $T_c$ for Lead. \emph{J. Low Temp. Phys.} \textbf{32}, 495-503 (1978).
\bibitem{R(p)/R(0)}
Eiling, A. \& Schilling, J. S. Pressure and temperature dependence of electrical resistivity of Pb and Sn from 1-300 K and 0-10 GPa-use as continuous resistive pressure monitor accurate over wide temperature range; superconductivity under pressure in Pb, Sn and In. \emph{J. Phys. F: Metal Phys.} \textbf{11}, 623-639 (1981).



\end{thebibliography}
\end{document}